# Observation of Yu-Shiba-Rusinov states in superconducting graphene

E. Cortés-del Río [1,2], J.L. Lado [3], V. Cherkez [4,5], P. Mallet [4,5], , J-Y. Veuillen [4,5], J.C. Cuevas [6, 2], J.M. Gómez-Rodríguez [1,2,7], J. Fernández-Rossier [8,9], I. Brihuega [1,2,7] *

[1] Departamento Física de la Materia Condensada, Universidad Autónoma de Madrid, E-28049 Madrid, Spain.

[2] Condensed Matter Physics Center (IFIMAC), Universidad Autónoma de Madrid, E-28049 Madrid, Spain.

[3] Department of Applied Physics, Aalto University, Espoo, FI-00076, Finland.

[4] Université Grenoble Alpes, CNRS, Institut Néel, F-38400 Grenoble, France.

[5] CNRS, Institut Neel, F-38042 Grenoble, France.

[6] Departamento Física Teórica de la Materia Condensada, Universidad Autónoma de Madrid, E-28049 Madrid, Spain.

[7] Instituto Nicolás Cabrera, Universidad Autónoma de Madrid, E-28049 Madrid, Spain

[8] QuantaLab, International Iberian Nanotechnology Laboratory (INL), Avenida Mestre José Veiga, 4715-310 Braga, Portugal

[9] Departamento de Física Aplicada, Universidad de Alicante, San Vicente del Raspeig 03690, Spain

**When magnetic atoms are inserted inside a superconductor, the superconducting order is locally depleted as a result of the antagonistic nature of magnetism and superconductivity[1]. Thereby, distinctive spectral features, known as Yu-Shiba-Rusinov states, appear inside the superconducting gap[2-4]. The search for Yu-Shiba-Rusinov states in different materials is intense, as they can be used as building blocks to promote Majorana modes[5] suitable for topological quantum computing[6]. Here we report the first realization of Yu-Shiba-Rusinov states in graphene, a non-superconducting 2D material, and without the participation of magnetic atoms. We induce superconductivity in graphene by proximity effect[7-9] brought by adsorbing nanometer scale superconducting Pb islands. Using scanning tunneling microscopy and spectroscopy we measure the superconducting proximity gap in graphene and we visualize Yu-Shiba-Rusinov states in graphene grain boundaries. Our results reveal the very special nature of those Yu-Shiba-Rusinov states, which extends more than 20 nm away from the grain boundaries. These observations provide the long sought experimental confirmation that graphene grain boundaries host local magnetic moments[10-14] and constitute the first observation of Yu-Shiba-Rusinov states in a chemically pure system.**

Superconducting (SC) order arises in many materials because of the formation of a coherent many-body state of electrons pairs with zero spin[15]. The addition or removal of one electron from this state, resulting in an unpaired electron, costs a small energy gap associated to the overhead of breaking a pair. Interactions that promote magnetism tend to destroy superconducting order in most materials. When perturbed with magnetic impurities, the superconducting order is depleted locally, which can give rise to the appearance of the in-gap bound states known as Yu-Shiba-Rusinov (YSR) states[2-4].

The measurement of YSR states on Mn adatoms on a superconductor, by means of scanning tunneling microscopy and spectroscopy (STM/STS), provided[16] the first direct observation of magnetism on an individual atom, as Anderson's theorem[1] precludes the existence of in-gap states if time reversal symmetry is preserved. In-gap YSR states have now been observed in a variety of systems that integrate superconductors with magnetic moments in the form of molecules[17], self-assembled[5] and artificial atomic chains[18], magnetic islands[19,20] and also in proximity-induced superconducting molecular break junctions[21]. Current interest in YSR states is driven by their potential to induce a topological phase transition in the superconductor[5,22], leading to the appearance of Majorana modes with potential application in topological quantum computing[6].

A priori, chemically pure graphene is far from being an optimal system to look for YSR states, as it lacks both superconductivity and magnetism. However, superconductivity can be induced in graphene via proximity effect[7-9] and magnetism is expected to occur in graphene grain boundaries (GBs)[10-14]. YSR states were predicted [23] to occur in graphene endowed with a superconducting proximity gap when a local moment is induced by chemisorption of individual hydrogen atoms, which are known to lead to the emergence of magnetic moments[24]. Given the ubiquity of graphene GBs, it is of the utmost importance to obtain unequivocal experimental evidence of the presence of local moments. Here, we assess this matter by inducing superconductivity in the GB via proximity effect and exploring the emergence of YSR states by means of STS spectroscopy. By so doing, we demonstrate that magnetism can coexist with proximity induced superconductivity, realizing exotic electronic phases in carbon-only structures, which complements the ongoing efforts along this line using twisted bilayer graphene[25,26].

We grow several layers of graphene on a SiC(000-1) substrate. In this system the rotational disorder of the graphene layers electronically decouples $\pi$ bands leading to a stacking of essentially isolated graphene sheets[27-29]. The graphene layer on the surface is neutral, with 100-500 nm wide single-crystal domains of different crystallographic orientations[30], see methods and Extended data Fig.1. On this graphene surface, we deposit 5-15 monolayers (ML) of Pb at a rate of 3-10 ML/min while maintaining the sample at room temperature[31]. As a result, several triangular Pb islands with heights between 2-10 nm and sides between 20-300 nm are formed, see Fig. 1a.

We carry out STS of the electronic properties of graphene in the vicinity of the Pb islands to measure the strength of the superconducting proximity coupling. We acquire conductance *dI/dV* curves, which probe the energy-resolved local density of states (LDOS(E)) under the tip position, as a function of the distance to the Pb islands. Our base experimental temperatures are: $T_{sample}=4K$, way below the Pb superconducting $T_c=7.2K$, and $T_{tip}=3K$. To increase the energy resolution beyond the thermal limit, we acquire STS spectra using SC Pb tips[22,32]. The conductance map, $dI/dV(x, E)$, plotted with respect to distance to the Pb island and energy in Fig. 1d, shows that a slowly decaying SC gap is induced in the pristine area of graphene, extending several tens of nm away from the Pb island edge, see also $dI/dV$ spectra in Fig. 1c and Extended Data Fig. 2.

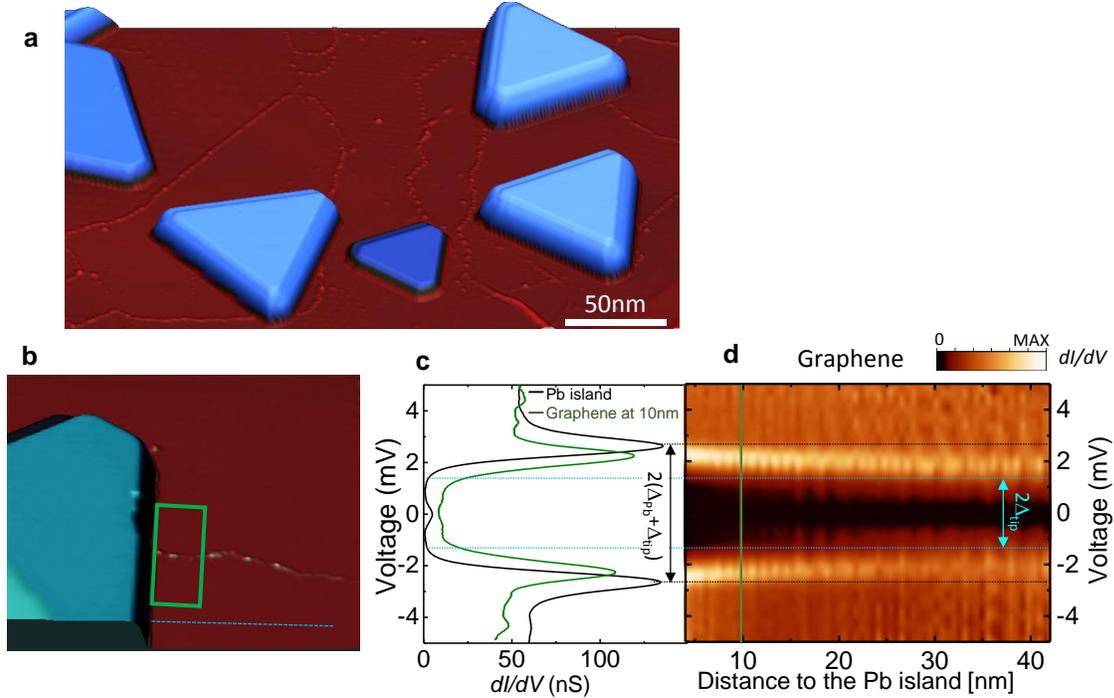

**Fig. 1 | Graphene superconductivity induced by Pb islands. a,** Large scale STM image showing the general morphology of the sample after Pb deposition. Triangular shaped Pb island are formed on top of the graphene surface grown on SiC(000-1). In the image, several GBs are also be observed. **b**, STM image showing a Pb island on a graphene region with a GB. **c**, $dI/dV$ spectra measured on the Pb island (black line) and on the pristine graphene region at 10 nm from it (green line), $V_{bias}$=10mV; $I_{set}$=0.5nA. **d**, Conductance map $[dI/dV(x,E)]$ along the dashed line in (B) showing how SC is induced in pristine graphene close to the Pb island and far away from the GB ($V_{bias}$=10mV; $I_{set}$=0.5nA). Horizontal dotted lines indicate the Pb island SC gap, $\Delta_{Pb}$ (black line) and the SC tip gap $\Delta_{tip}$, (blue line). Vertical dotted line correspond to the graphene spectrum of panel C). STS data are measured with a SC tip at $T_s$=4K; $T_{tip}$=3K. All the STM data were measured using the WSxM software[33].

We now turn our attention to the STS on the GB of our superconducting graphene. Graphene GBs are naturally formed at the frontiers where the different graphene domains meet. These boundary regions present a rich structure characterized by the presence of many under-coordinated carbon atoms[11,13,14,30], usually seen as bright features in STM images (Fig. 1a,b, 2a, and Extended Data Figs. 1, 3). An atomically resolved STM image of the graphene GB, highlighted by a green rectangle in Fig. 1b, is shown in Fig. 2a. The GB horizontally crosses the middle of the image, separating two graphene grains with 25º atomic lattice misalignment

[see Fig. 2b and Extended Data Fig. 2]. A 3.2 nm height Pb island, perpendicularly crossing the GB on the left side, induces SC in the graphene region, see Fig. 1b-d.

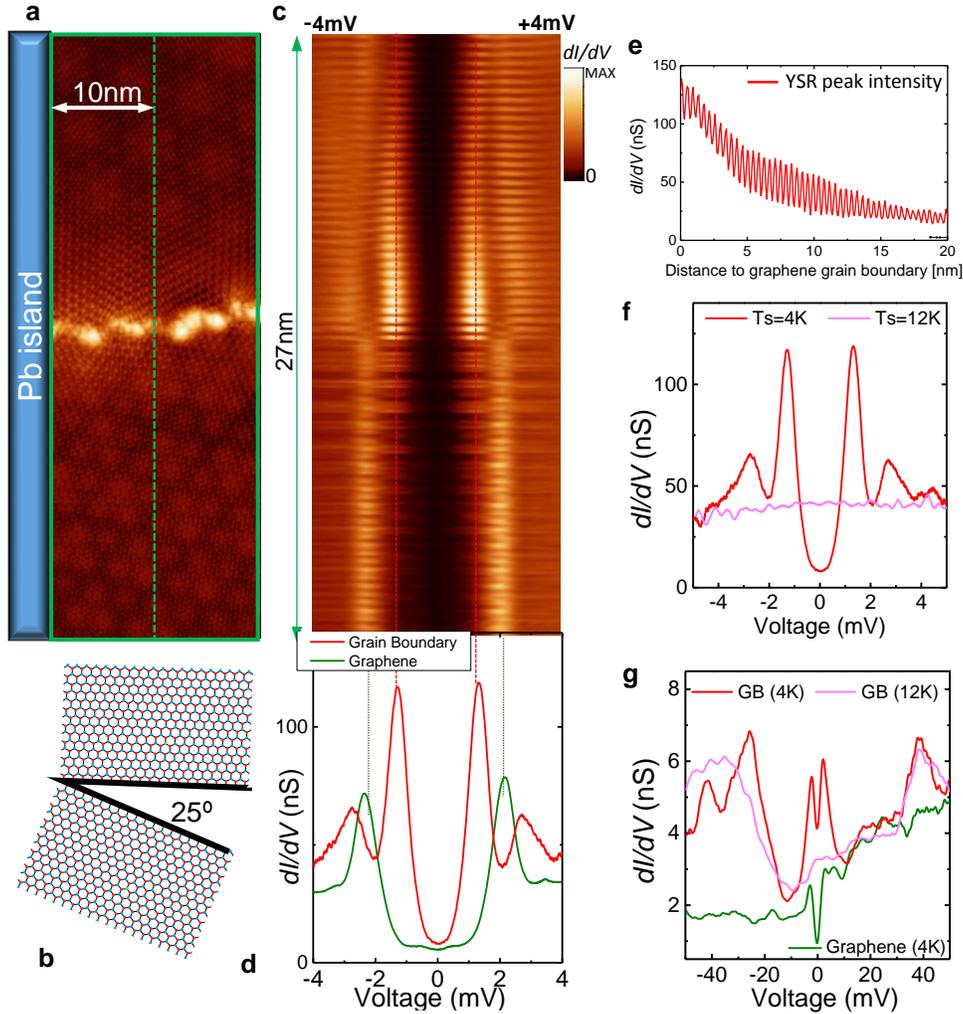

**Fig. 2 | GB magnetism induces Yu-Shiba-Rusinov states in superconducting graphene.**
**a**, Atomically resolved topography showing a zoom in of the GB outlined by a green rectangle in Fig. 1b. Superconductivity is induced in the region by a Pb island placed just at the left edge of the image. **b**, Schematic of the orientation of the graphene domains in (A), showing the 25° rotation between them. **c**, Conductance map [$dI/dV(x,E)$], measured with a SC tip, along the line crossing the GB, highlighted in green in A) ($V_{bias}$=12mV; $I_{set}$=0.5nA). The line is parallel to the $\sqrt{3} \times \sqrt{3}$ direction of the upper graphene grain and to the Pb island (at a constant distance of 10nm). The vertical dotted lines outline the YSR in-gap states. **d**, Single $dI/dV$ spectra on top of the GB (red line) and on pristine graphene (green line), both measured at a 10nm distance from the Pb island ($V_{bias}$=12mV; $I_{set}$=0.5nA). **e**, Spatial extension of the in-gap YSR state and $\sqrt{3} \times \sqrt{3}$ modulation. The graph corresponds to a measurement of the $dI/dV$ YSR peak intensity as a function of the distance to the GB, on the upper graphene grain. **f**, $dI/dV$ spectra measured on the same spot of the GB for temperatures both below (red) and above (pink) the superconducting transition ($V_{bias}$=12mV; $I_{set}$=0.5nA).

YSR in-gap state completely vanishes for $T>T_c$ (pink curve). **g**, $dI/dV$ spectra on a higher energy scale ($V_{bias}$=50mV; $I_{set}$=0.25nA). On the GB (red and pink curves), besides the YSR state, two broader peaks at higher energies are also observed. Those peaks, ascribed to GB magnetism, are essentially unaltered through the SC phase transition.

Our STS $dI/dV$ spectra (Fig. 2c), acquired moving the STM tip in the direction perpendicular to the graphene GB at a constant distance of 10 nm from the Pb island -along the green dotted line in Fig. 2a- show the unambiguous presence of in-gap YSR states in the graphene GB (see also point spectra in Fig. 2d). The evolution of the $dI/dV$ spectra as the STM moves perpendicular to the graphene GB, show how the amplitude of the YSR features is maximal at the graphene GB and decays as the tip enters into the pristine graphene areas. The extension of the YSR feature in that direction is at least 20 nm (Fig. 2e).

The spatial extension of these YSR states is much larger than the one reported by Menard et al in the case of magnetic dopants in two dimensional NbSe$_2$[34] that in turn is much larger than the ones reported in 3D superconductors. Moreover, the YSR state features a $\sqrt{3}\times\sqrt{3}$ modulation along the direction rotated 30º with respect to the atomic lattice, see Fig. 2e. Such modulation, associated to the wave vector which spans the two valleys in the Brillouin zone, is also observed in the resonance magnetic states generated close to the Fermi level ($E_F$) by atomically sharp impurities in graphene[24], and reflects in both cases the sublattice dependent response of graphene to local perturbations.

Further experimental evidence of the magnetic origin of the in-gap features observed at the GB can be obtained from STS carried out by increasing the temperature above the superconducting $T_c$, so that superconducting order is suppressed. This brings graphene to its normal phase. In this situation, both the gap and the sharp in-gap state completely disappear (see Fig. 2f). In addition, we have measured temperature dependent spectra on a larger voltage scale (see Fig. 2g). Our $dI/dV$ spectra on the graphene GB show the presence of two additional broad peaks, one below and one above $E_F$, at energies well beyond the SC gap. We ascribe those peaks to the addition/removal quasiparticle peaks in localized states, whose splitting comes mostly from the Coulomb repulsion[24,35,36]. Consistent with this assumption, our data show that, contrary to the in-gap states, such peaks are essentially independent to temperature changes throughout graphene SC transition, see Fig. 2g.

In order to determine the actual density of states (DOS), we have to carry out a numerical deconvolution of the $dI/dV$ curves to remove the features that arise from the tip superconducting gap, see methods for details. The deconvoluted curves so obtained provide a faithful representation of the surface DOS having performed the measurements with a superconducting tip, and permit to address the energy of the YSR states inside the gap.

The hallmark of YSR states are pairs of peaks with energy $E_B<\Delta$ symmetrically located around $E_F=0$, but with asymmetric height, on account of their different electron-hole weight. In the case of the GB shown in Fig. 2, the YSR state correspond to a state with $E_B\approx0$, as shown in the deconvoluted spectra of the middle panel of Fig. 3b, see also Extended Data

Fig.4. Different spectra, with $E_B \neq 0$, are obtained, depending on the GB-Pb island configuration and the location of the boundary probed, see Fig. 3 and Extended Data Fig. 5. The variation in energy of the bound state suggests that the strength of the interaction induced exchange coupling depends on the actual details of the interface.

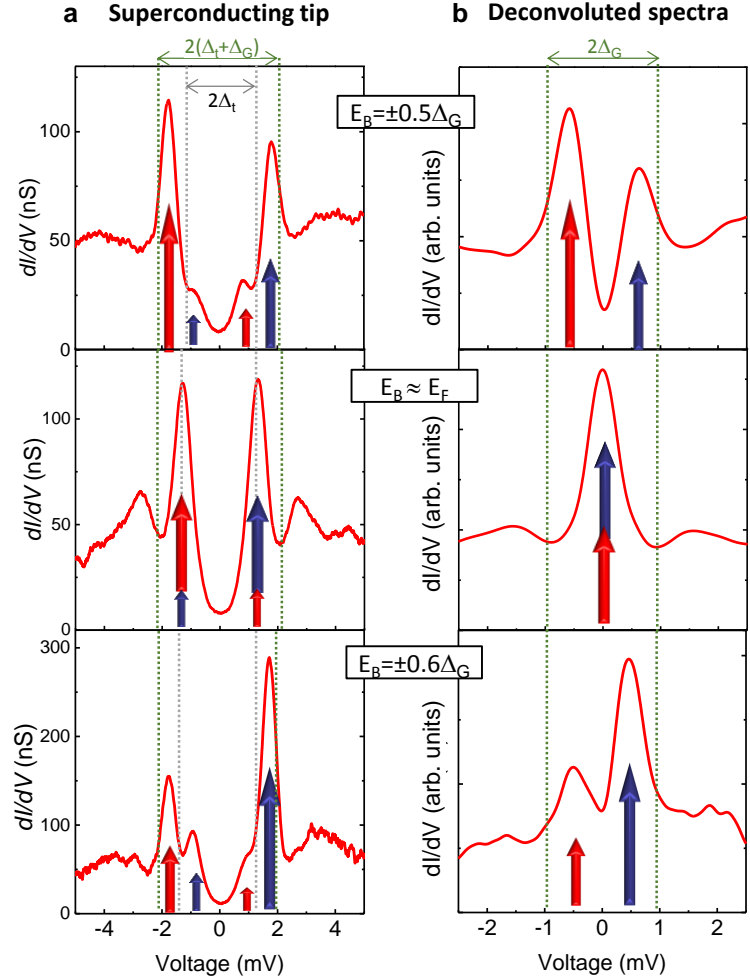

**Fig. 3 | Yu-Shiba-Rusinov states in graphene for different GB configurations**. **a**, $dI/dV$ spectra measured, with a SC tip, on 3 characteristic GB configurations (stabilization values for the up, middle and bottom spectra are $V_{bias}$=10mV; $I_{set}$=0.5nA; $V_{bias}$=12mV; $I_{set}$=0.5nA; $V_{bias}$=6mV; $I_{set}$=0.5nA respectively). Outer vertical dotted lines outline the energy position of $\Delta_{tip}+\Delta_G$. Inner vertical dotted lines outline the energy position of $\Delta_{tip}$. In the spectra, 4 in-gap peaks, outlined by red and blue arrows, are clearly resolved. The 2 outer ones, which larger amplitudes, correspond to the electron and hole-like components of the YSR excitation. The 2 inner ones, at $E < \Delta_{tip}$ and with much smaller amplitudes, are due to the thermally activated tunneling occurring at finite temperatures. **b**, Corresponding GB $dI/dV$ spectra, obtained by numerical deconvolution of STS data in panel a. Vertical dotted lines outline the energy position of $\Delta_G$.

In general, the number of YSR states, and their binding energies, depend both on the internal structure of the magnetic impurity and on the strength of the exchange coupling *J* between the local spin and the surface electrons. Thereby, in other systems, for the same magnetic species placed in different adsorption sites of the surface of a conventional SC, different YSR spectra were also reported [17,22]. Depending on the strength of *J*, the parity of the ground state of the system can be different. In the weak exchange coupling limit, the BCS condensate remains in the spin *S=0* limit, and the multiplicity of the state is given by the one of the local moment. However, at some critical value of *J* a Cooper pair is broken, so that the BCS condensate has *S=1/2*. The critical point is marked by a YSR with $E_B=0$ [37,38]

The nature of the YSR states reported here is peculiar on several counts. First, the YSR states are observed in a chemically homogeneous region, made of carbon only and free of other chemical species. Second, they are observed in graphene, that has to borrow superconductivity from Pb islands located several nanometers away. Third, YSR states present an extremely large spatial extension, on account of the 2D nature of graphene [34].

In order to provide a theoretical basis to this completely novel scenario, we model the GB using a Hubbard model plus a pairing BCS term $\mathcal{H} = \sum_{i,i',\sigma} t_{ii'} c_{i\sigma}^\dagger c_{i'\sigma} + U \sum_i n_{i\uparrow} n_{i\downarrow} + \Delta \sum_i c_{i\uparrow}^\dagger n c_{i\downarrow}^\dagger + h.c.$ We consider a GB between a zigzag and an armchair oriented ribbons, Fig. 4a. The mismatch angle is 30º, close to the one observed experimentally. Taking *U=Δ=0*, the single particle bands feature a narrow band associated to states located at the GB, see Fig. 4b. When a finite value of *U=2t* is considered, in the mean field approximation, still with *Δ=0*, we find the emergence of a spin-splitting in the band structure (Fig. 4f) associated to magnetism in the GB, as shown in Fig. 4e. Let us now move to the case with a proximity induced superconducting pairing stemming from the Pb islands, focusing first on the non-magnetic situation. In this case, in which we ignore electronic interactions, the interface remains non-magnetic giving rise to a superconducting gap everywhere as shown in Fig. 4c. In stark contrast, when we consider the magnetic case with superconducting proximity effect (Fig. 4g), we observe the emergence of the in-gap Yu-Shiba-Rusinov states inside the superconducting gap.

It is also informative to look at the density of states as a function to the distance to the GB, as shown in Figs. 4d,h. In the presence of superconductivity, in-gap resonances at the GB appears when the interaction-induced magnetism is included (Fig. 4h), whereas no in-gap resonance appears in the absence of magnetism (Fig. 4d).

There are several features to highlight. First, the YSR peaks show an electron-hole asymmetry as observed experimentally, which can be related to the breakdown of the bipartite character of the GB, that leads to single-particle states without electron-hole symmetry. Second, the location in energy of the in-gap states depends on details of the interface, such as the strength of the interaction-induced exchange coupling, or the specific geometry of the GB, see methods and Extended Data Figs. 6-8 for details. We note that our calculations do not incorporate the effect of chemical reconstructions, but rather focus on the fundamental mechanism in which an interface-defect gives rise to YSR states. Finally, the emergence of flat bands and magnetism orientation interfaces are generic features of GBs, beyond the specific case considered here. This stems from the Berry phase mismatch between the orientations, that enforces the emergence of flat band states for a generic chiral interface[39].

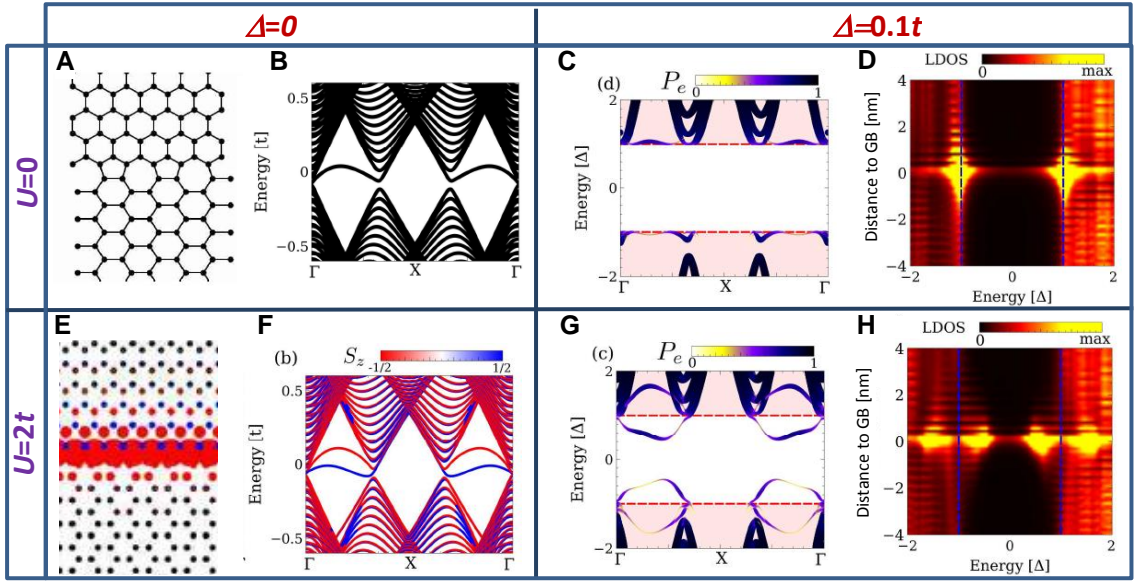

**Fig. 4 | Interface between an armchair and zigzag oriented regions**. **a**, Atomic structure of the GB. Such GB hosts a nearly flat band as shown in **b**. When superconducting proximity effect is introduced, in the absence of magnetism, no in-gap state is observed as shown in the band structure, **c,** and in the spatially resolved DOS **d**. Upon inclusion of interactions, the GB becomes magnetic, with the appearance of local magnetic moments, **e**, and the spin-splitting of the band structure, **f**. The area of each circle in e is proportional to the amplitude of the local magnetic moment at each atom. When superconducting proximity effect is introduced on top of the magnetic state, in-gap YSR bands appear, **g, h**. $P_e$ denotes the projection of the state in the electron-sector of the Bogoliubov-de Gennes Hamiltonian, which are the states accessible in STM spectroscopy.

To summarize, we have used Pb islands to induce superconductivity in graphene, exploiting GB as a source of local magnetic moments to realize, for the first time, YSR states in graphene. Importantly, our experiments provide an unequivocal proof of carbon magnetism at graphene GBs, as the emergence of in-gap state in the presence of superconductivity demonstrate the interaction with local magnetic moments. Finally, our results provide a starting point towards the exploration of exotic electronic phases involving both magnetism and superconductivity in graphene that, together with further proximity to strong spin-orbit coupling materials, may lead to the emergence of topological superconductivity at graphene GBs.

**Acknowledgments:** This work was supported by AEI and FEDER under projects MAT2016-80907-P and MAT2016-77852-C2-2-R (AEI/FEDER, UE), by the Fundación Ramón Areces, and by the Comunidad de Madrid NMAT2D-CM program under grant

S2018/NMT-4511. J. F. R. acknowledges financial support European Regional Development Fund Project No. NORTE-01-50145- FEDER-000019, and the UTAPEXPL/NTec/0046/2017 projects, as well as Generalitat Valenciana funding Prometeo2017/139 and MINECO Spain (Grant No. MAT2016-78625-C2). J.L.L is grateful for financial support from the Academy of Finland Projects No. 331342 and No. 336243.

**Author contributions:**

**Additional information:**

**Supplementary Information** is available for this paper

**Materials and Correspondence**

* Correspondence and requests for materials should be addressed to:
ivan.brihuega@uam.es

**Data availability:** The data that support the findings of this study are available from the corresponding author upon reasonable request.

# METHODS

**Sample preparation and experimental details.**

All the preparation procedures and measurements were performed under UHV conditions. During the whole process -imaging pristine graphene sample => depositing Pb on it => and imaging it back- the sample was maintain in the same UHV system.

The multilayer graphene substrate was grown on a 6H-SiC(000-1) sample (C face) following the method described in [40]. This process takes place in a RF furnace. In brief, it consists in heating (at 1600°C) the SiC substrate held in a graphite crucible under an Ar atmosphere (1 bar Ar) for 30 minutes. Before this graphitization step, the substrate is etched in an Ar/H2 mixture[41]. After the growth, the sample is transferred into a separate ultrahigh vacuum (UHV) setup and outgazed.

The graphene layer on the surface has 100-500 nm wide single-crystal domains of different crystallographic orientations [30]. On the surface, graphene GBs are naturally formed at the frontier between these pristine graphene domains with different crystallographic orientations. These boundary regions are usually seen as bright features in STM images (see GB outlined by arrows in Fig. E1).

In this system, the rotational disorder of the graphene layers electronically decouples π bands leading to a stacking of essentially isolated graphene sheets [27-29]. The pristine graphene surface layer is essentially neutral, presenting a very low electron doping [31] ($<3\times10^{11}$ cm$^{-2}$) as deduced from the quasiparticle interference (QPI) pattern measurements [8,9] and from the position of the Van Hove singularities associated with the moiré pattern [29, 44]. On this graphene surface, we deposit 5-15 monolayers (ML) of Pb at a rate of 3-10 ML/min while maintaining the sample at RT [31]. We obtained the thickness of the islands in monolayer (ML) units by dividing their apparent height as measured by STM by the Pb (111) interlayer distance (0.286 nm). As a result, several triangular Pb islands with heights between 2-10 nm and sides between 20-300 nm are formed, see Extended Data Fig. 1.

**Superconductivity induced in graphene by Pb islands.**

We induce superconductivity in graphene by the proximity coupling with adsorbed nanometer sized Pb islands. To measure the strength of the superconducting proximity coupling, we probe by STS the electronic properties of pristine graphene in the vicinity of the Pb islands. We show an example of such measurement in Extended Data Fig. 2, where we present the evolution of the SC gap induced in graphene as a function of the distance to a Pb island. To increase the energy resolution beyond the thermal limit [32,22], we acquire STS data using SC Pb tips, which in the present work are obtained by crashing a Pt/Ir tip on Pb.

Extended Data Fig. 2c shows a *dI/dV(x, E)* map, measured with a SC Pb tip, as a function of the distance to the Pb island and energy. Then, as described in the next section, we numerically deconvolute the *dI/dV(E)* curves to obtain a faithful representation of the surface DOS, see Extended Data Fig. 2d. In addition, in Fig. S1e, we show a *dI/dV* map, measured using a normal (non-SC) tip along exactly the same graphene line, which allow us to confirm the validity of our deconvolution method. In all cases -SC tip, deconvoluted data, normal tip- the conductance *dI/dV* maps, -Extended Data Figs. 2c, 2d and 2e, respectively- shows a slowly decreasing SC gap induced in the pristine area of graphene, which extends several tens of nm away from the Pb island edge. In Extended Data Fig. 2f, h we show single *dI/dV* spectra measured, on the Pb island, with the same SC tip (Extended Data Fig. 2f) and normal tip (E2H) as the *dI/dV* maps plotted bellow. Extended Data Fig. 2g show the DOS of the tip obtained from the numerical deconvolution of the experimental data measured with the Pb SC tip (Extended Data Fig. 2c, f).

**Method for the numerical deconvolution of the experimental *dI/dV* data**

In order to increase the energy resolution beyond the thermal limit [11,12], we acquire STS data using SC Pb tips. Therefore, to determine the actual density of states (DOS), we have to carry out a numerical deconvolution of the *dI/dV* curves to remove the features that arise from the tip superconducting gap. The deconvolution is performed by defining a functional $S = \int_{-\infty}^{\infty} dw [g_{th}(w) - g_{exp}(w)]^2$, where $g_{exp}(w)$ is the experimentally measured *dI/dV* and $g_{th}(w)$ is the theoretically expected *dI/dV* of the form $g_{th}(w) = \frac{\partial}{\partial w} \int dw' D_{tip}(w + w') D_{surface}(w')[1 - n_{tip}(w + w')] n_{surface}(w)$, where $D_{tip}$ is the density of states of the tip (which is known) and $D_{surface}$ the density of states of the surface. The density of states of the surface is computed by minimizing the functional $S$ as $\frac{\delta S}{\delta D_{surface}} = 0$. The deconvoluted curves so obtained provide a faithful representation of the surface DOS having performed the measurements with a superconducting tip, and permit to address the energy of the YSR states inside the gap.

**Calculations for different graphene superconducting GB configurations.**

Our theoretical model show that the location in energy of the in-gap states depends on details of the interface, such as the strength of the superconducting proximity coupling, the strength of the interaction-induced exchange coupling, or the specific geometry of the GB. To illustrate this, we show in Extended Data Figs. 6 and 7 the results of the calculations made for the same GB geometry as in Fig. 4 of the main manuscript, for different values of the graphene SC induced gap *Δ* (Extended Data Fig. 6), and of the exchange coupling *J*, effectively controlled by *U* (Extended Data Fig. 7). In addition, in Extended Data Fig. 8, we show the results for a different GB geometry, also presenting 30º rotation between the domains. In this particular case, as in the case of Fig. 4 of the main manuscript, we have chosen *U = 2t* and *Δ = 0.1t*.

Our calculations show the robustness of our results, with the emergence of YSR states in graphene for various GB configurations, when interactions, and thus magnetism, are included. In the absence of interactions, a nearly flat band is located at the interface (Fig. 4e and Extended Data Fig. 8b), that becomes spin-split when interactions are turned on (Fig. 4f and Extended Data Fig. 8c) giving rise to YSR states when SC is included (Fig. 4h and Extended Data Fig. 8h).

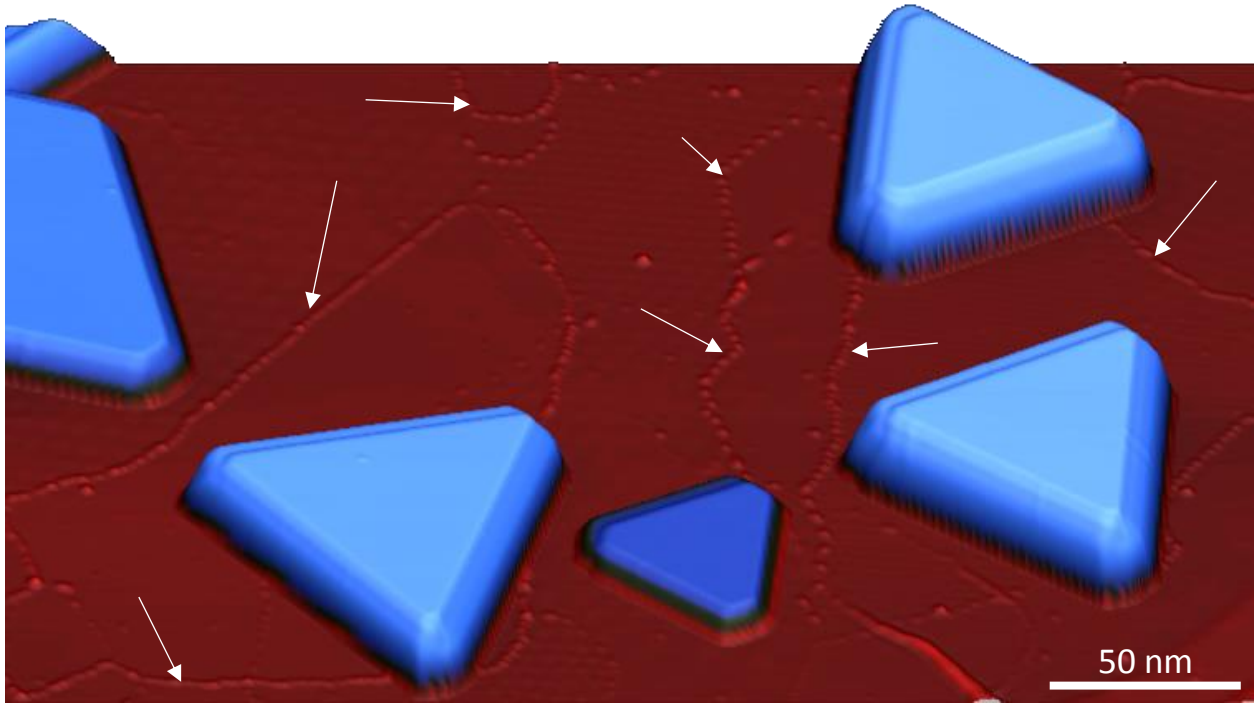

**Extended Data Fig.1 | Sample topography.** STM data showing the same image as in Fig. 1a of the main manuscript represented at larger size. The STM image show the general topography of the sample with triangular Pb islands on the graphene surface. Grain boundaries, outlined by white arrows, are seen in STM images as bright linear defects. $V_{bias}$=1.0V; $I_t$ = 0.05Na

.

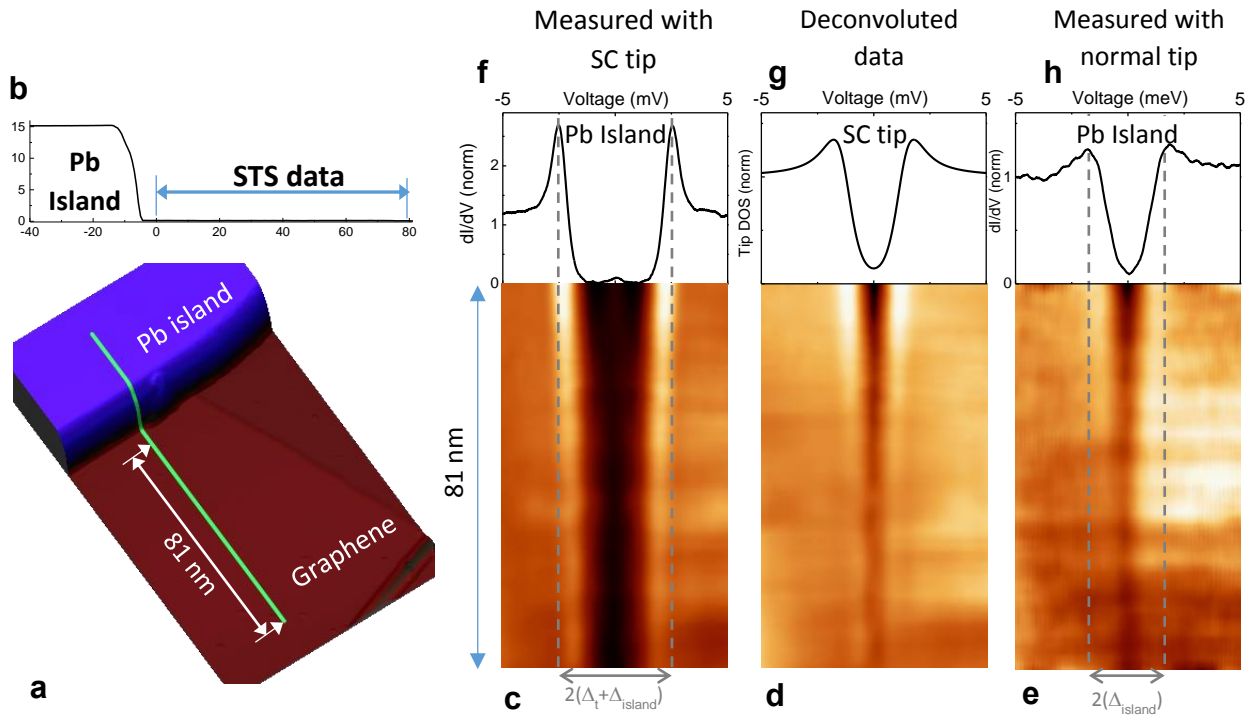

**Extended Data Fig. 2 | Superconductivity induced on graphene by a Pb island**. **a**, STM topography of a graphene region (plotted on red colors) with a Pb island (plotted on blue colors) on top. **b**, Line profile corresponding to the green line outlined in **a**. **c**-**e**, Conductance maps [$dI/dV(x,E)$], measured as a function of distance to the Pb island along the section of the green line outlined by arrows in (**a**, **b**). The $dI/dV(x,E)$ maps correspond to data measured with a SC tip ,**c**, its corresponding numerical deconvolution with the method detailed in SOM4 ,**d**, and to data measured, along exactly the same graphene line, with a normal, non-SC, tip ,**e**. **f**, **h**, Single $dI/dV$ spectra measured on top of the Pb island with the same SC tip , **f**, and normal tip , **h**, as the $dI/dV$ maps bellow. **g**, DOS of the tip extrated from the numerical deconvolution of the experimental data of panels **g** and **f**.

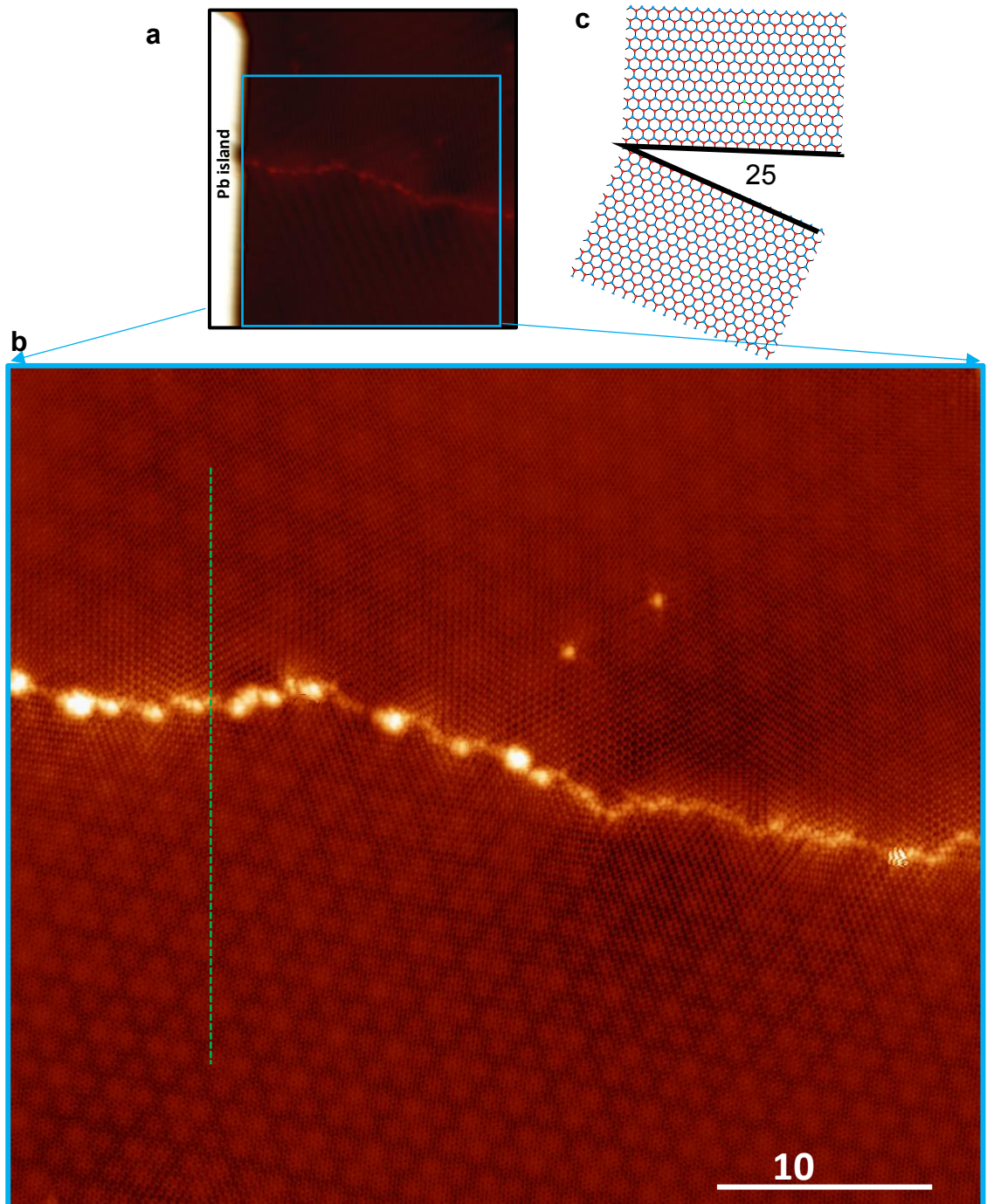

**Extended Data Fig. 3 | Topography of the GB shown in Figs 1 and 2 of the main manuscript. a**, 50x50nm$^2$ STM image showing a detail of the same region as in Fig 1a and 2b of the main manuscript where a Pb island on a graphene region with a grain boundary can be observed. **b**, Atomically resolved STM image showing a larger view of the GB shown in Fig 1b of the main manuscript. The green dotted line outlines the graphene region where the dI/dV map of Fig. 2 was measured. **c**, Schematic of the orientation of the two graphene domains, below and above the GB, showing the 25° rotation between them.

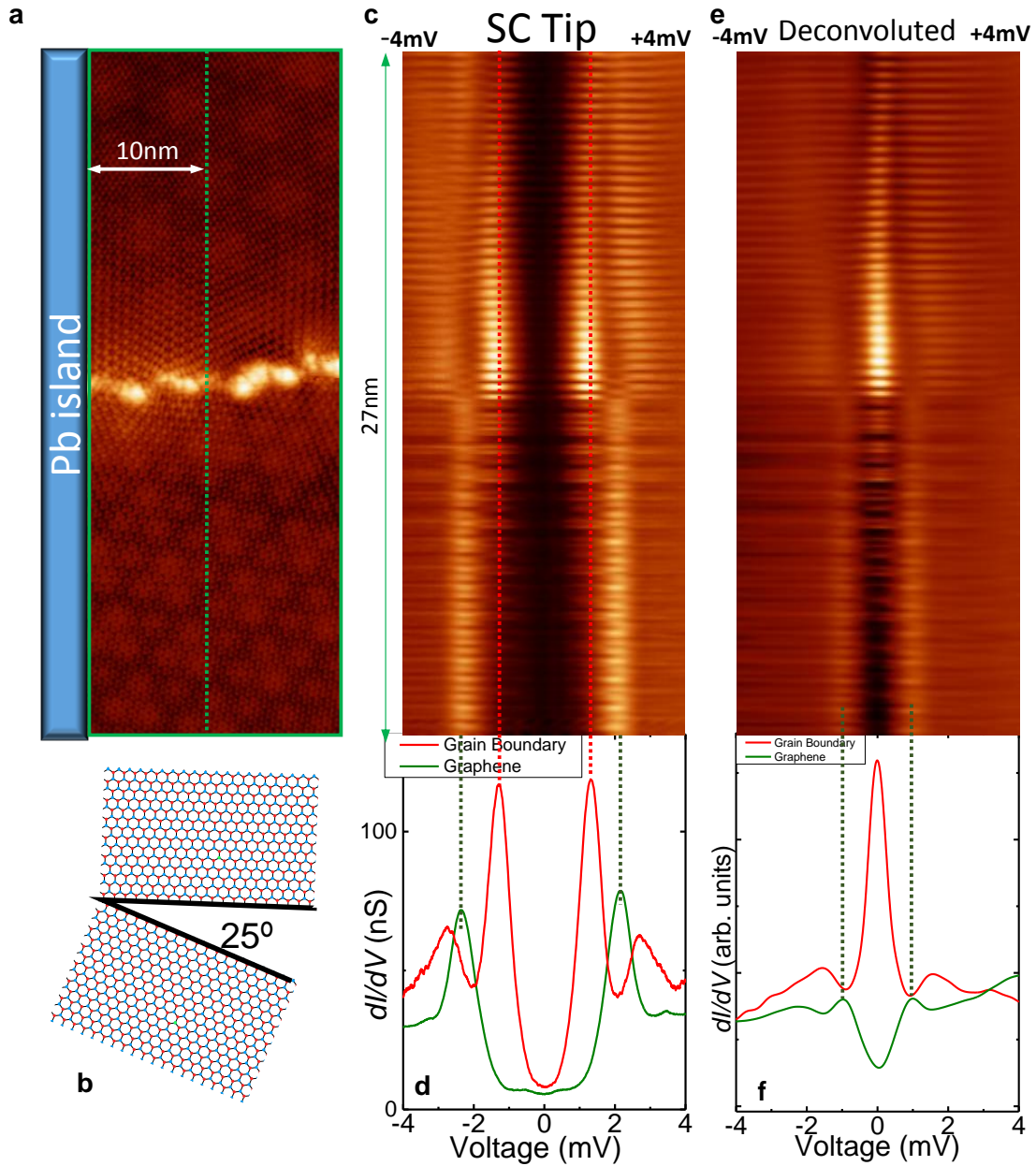

**Extended Data Fig. 4 | Numerical deconvolution of the STS data shown in Fig.2 of the main manuscript. a**, Atomically resolved topography showing a zoom in of the grain boundary shown in Figs. 1**a**,2, E2. Superconductivity is induced in the region by a Pb island placed just at the left edge of the image. **b**, Schematic of the orientation of the graphene domains in A), showing a 25º rotation between them. **c**, Conductance map [*dI/dV(x,E)*], measured with a SC tip, along the dotted line crossing the grain boundary, highlighted in green in a. The line is parallel to the R3 direction of the upper graphene grain and to the Pb island (at a constant distance of 10nm). The vertical dotted red lines outline the YSR In-gap states. **d**, Single *dI/dV* spectra on top of the GB (red line) and on pristine graphene (green line), both measured at a 10nm distance from the Pb island. **e**, **f**, STS data from panels **c** and **d** respectively, after numerical deconvolution.

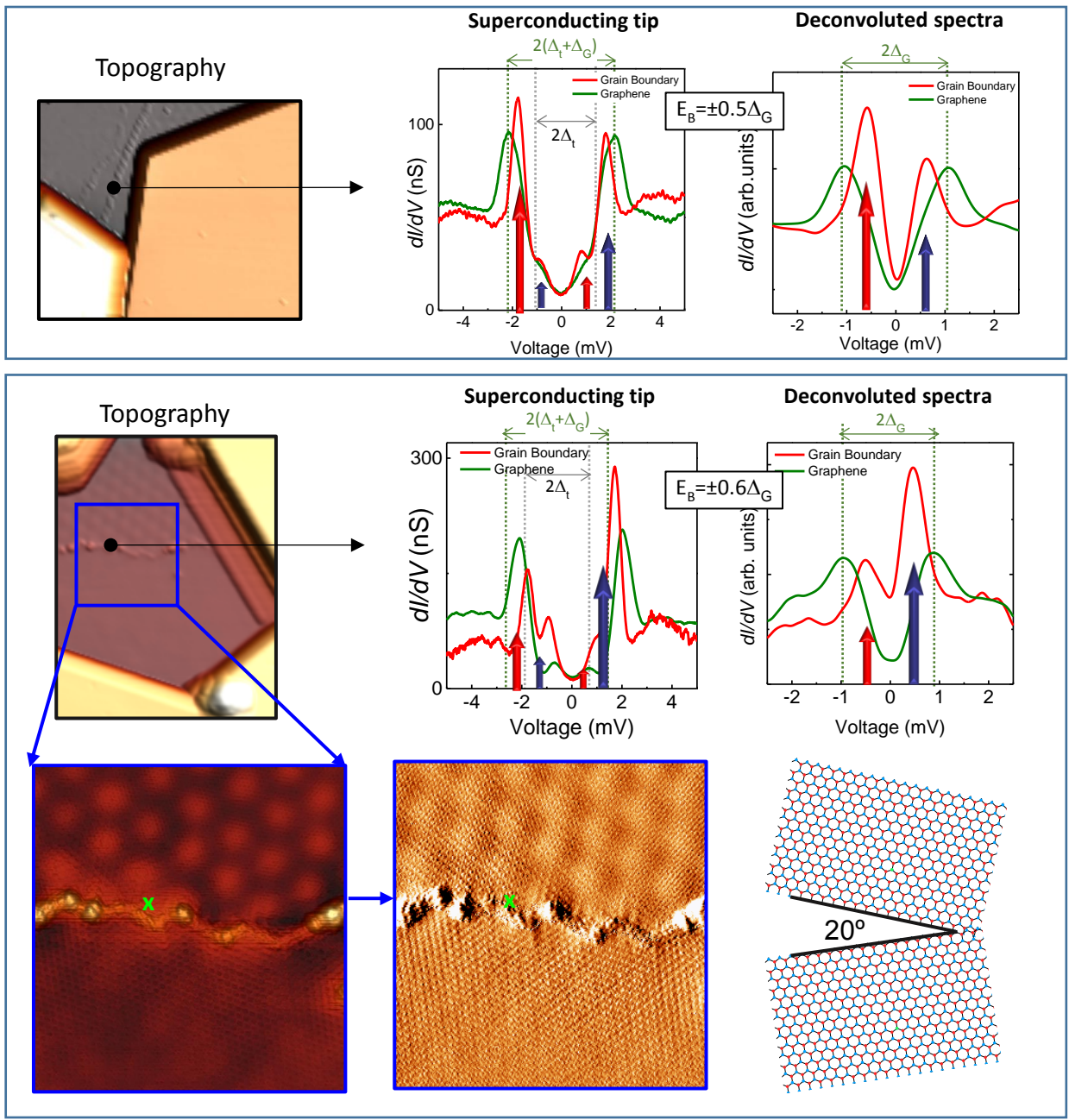

**Extended Data Fig. 5 | Data associated with Figure 3 of the main manuscript.** Top panel, data associated with STS measurements shown on upper panel in Fig 3 of the main manuscript. On the left, a topographic STM image of the region is shown. On the right, the raw and deconvoluted *dI/dV* spectra measured on the GB (red) and graphene (green) are shown. Lower panel, data associated with STS measurements shown on the lower panel in Fig. 3 of the main manuscript. The topography of the region is shown, together with a schematic of the orientation of the graphene domains, showing a 20º rotation between them. The raw and deconvoluted *dI/dV* spectra measured on the GB (red) and graphene (green) are also shown.

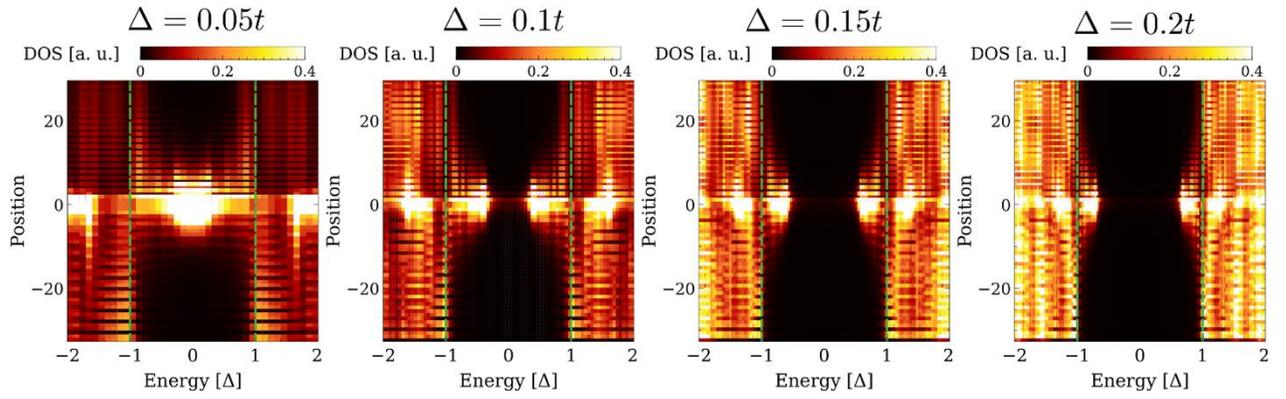

**Extended Data Fig. 6 | YSR states for different Δ.** Local density of states, for the same GB geometry as in Fig. 4 of the main manuscript, U = 2t and different values of the graphene induced gap Δ.

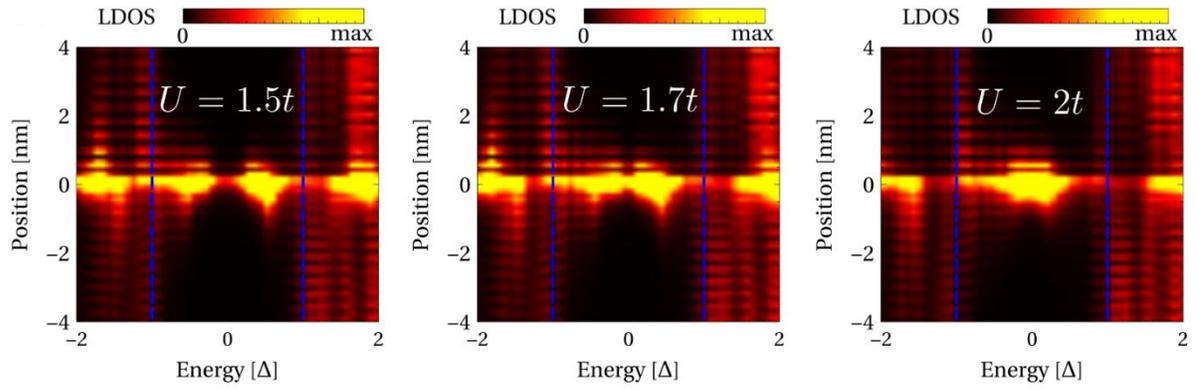

**Extended Data Fig. 7| YSR states for different J**. Local density of states, for the same GB geometry as in Fig. 4 of the main manuscript, $\Delta = 0.05t$ and different U values (that effectively controls J).

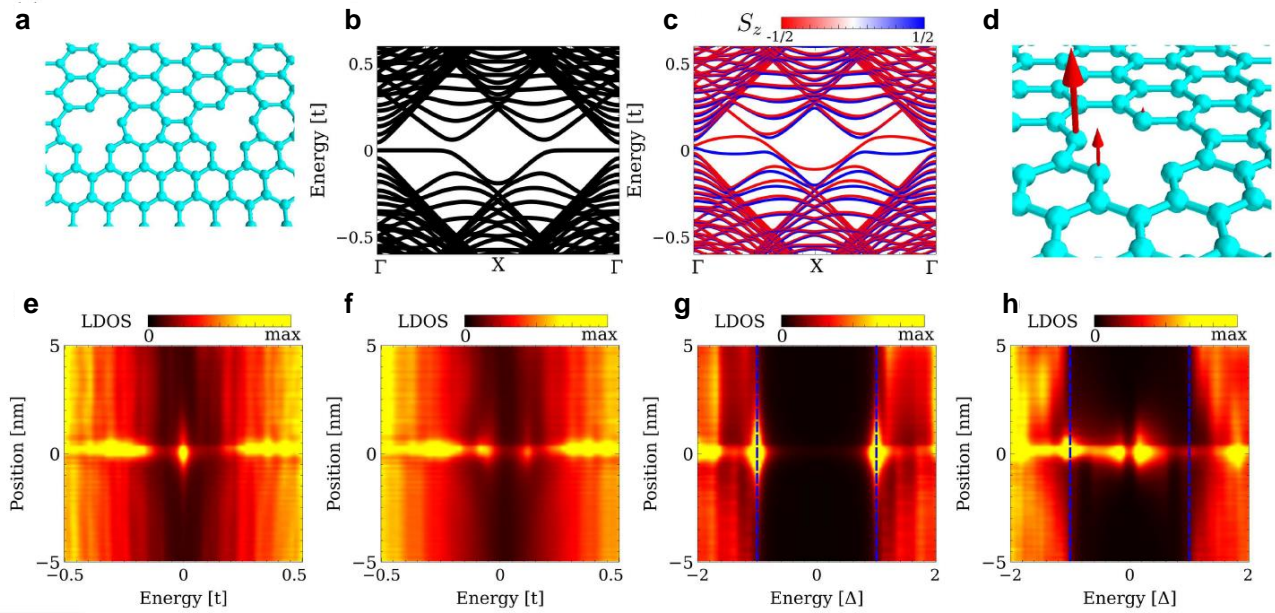

**Extended Data Fig. 8| YSR states for a different GB geometry**. Alternative geometry, with respect to the one shown in Fig. 4 of the main manuscript, for an interface between an armchair and zigzag oriented regions, showing an atomic defect, **a**. The band structure of such interface hosts a flat band, as depicted in, **b**. Upon introduction of electronic interactions, a spin splitting emerges in the selfconsistent band structure, **c**, giving rise to a local magnetic moment in the defect as shown in, **d**. Panels **e-h** show the local density of states, in the absence of magnetism and superconductivity, **e**, in the presence of magnetism and absence of superconductivity, **f**, in the presence of superconductivity and absence of magnetism, **g**, and in the presence of superconductivity and magnetism, **h**.